\documentstyle[12pt]{article}

\makeatletter

\newif\ifnoncomplete

\def\final{\noncompletefalse}
\noncompletefalse

\newif\ifrefphysrev

\refphysrevfalse

\def \vol(#1,#2,#3){\ifrefphysrev{{\bf {#1}},
{#3} (19{#2})}\else{{{\bf {#1}}(19{#2}){#3}}}\fi}
\def \NP(#1,#2,#3){Nucl.\ Phys.\          \vol(#1,#2,#3)}
\def \PL(#1,#2,#3){Phys.\ Lett.\          \vol(#1,#2,#3)}
\def \PRL(#1,#2,#3){Phys.\ Rev.\ Lett.\   \vol(#1,#2,#3)}
\def \PRp(#1,#2,#3){Phys.\ Rep.\          \vol(#1,#2,#3)}
\def \PR(#1,#2,#3){Phys.\ Rev.\           \vol(#1,#2,#3)}
\def \PTP(#1,#2,#3){Prog.\ Theor.\ Phys.\ \vol(#1,#2,#3)}
\def \ibid(#1,#2,#3){{\it ibid.}\         \vol(#1,#2,#3)}

\def\thebibliography#1{%\par\newpage
\section*{References\@mkboth
  {REFERENCES}{REFERENCES}}\list
  {[\arabic{enumi}]}{\setlength\labelwidth{2ex}%[#1]}%
   \setlength\labelsep{0.05in} % separation between [1] and the first name.
   \setlength\leftmargin{0.25in}
    \setlength\itemsep{0pt}
    \setlength\parsep{0pt}
   \itemsep\parskip
    \usecounter{enumi}}
    \def\newblock{\hskip .11em plus .33em minus -.22em}
    \sloppy
    \sfcode`\.=1000\relax}

\def\@bibitem#1{\item\if@filesw \immediate\write\@auxout
       {\string\bibcite{#1}{\the\c@enumi}}\fi\ignorespaces
       {\ifnoncomplete\reversemarginpar{\hspace*{-1.05in}\makebox[1in][l]
       {{\footnotesize{\sl [#1]}}}}\fi}%
       }

\def\@cite#1#2{\unskip\nobreak\relax
    {[#1]}} % PRINT FORM of citation

\def\citenum#1{{\def\@cite##1##2{##1}\cite{#1}}}
\def\citea#1{\@cite{#1}{}}

\newcount\@tempcntc
\def\@citex[#1]#2{\if@filesw\immediate\write\@auxout{\string\citation{#2}}\fi
  \@tempcnta\z@\@tempcntb\m@ne\def\@citea{}\@cite{\@for\@citeb:=#2\do
    {\@ifundefined
       {b@\@citeb}{\@citeo\@tempcntb\m@ne\@citea\def\@citea{,}{\bf ?}\@warning
       {Citation `\@citeb' on page \thepage \space undefined}}%
    {\setbox\z@\hbox{\global\@tempcntc0\csname b@\@citeb\endcsname\relax}%
     \ifnum\@tempcntc=\z@ \@citeo\@tempcntb\m@ne
       \@citea\def\@citea{,}\hbox{\csname b@\@citeb\endcsname}%
     \else
      \advance\@tempcntb\@ne
      \ifnum\@tempcntb=\@tempcntc
      \else\advance\@tempcntb\m@ne\@citeo
      \@tempcnta\@tempcntc\@tempcntb\@tempcntc\fi\fi}}\@citeo}{#1}}
\def\@citeo{\ifnum\@tempcnta>\@tempcntb\else\@citea\def\@citea{,}%
  \ifnum\@tempcnta=\@tempcntb\the\@tempcnta\else
   {\advance\@tempcnta\@ne\ifnum\@tempcnta=\@tempcntb \else \def\@citea{--}\fi
    \advance\@tempcnta\m@ne\the\@tempcnta\@citea\the\@tempcntb}\fi\fi}

\def\affiliation#1{\cr
\makebox[0in]{\parbox{8in}{\begin{center} {\sl #1}\end{center}}} \cr}
\def\@affiliation{}

\def\and{\cr \makebox[0in]{\rule[-1cm]{0mm}{1cm}and } \cr}

\def\maketitle{\par
 \begingroup
 \def\thefootnote{\fnsymbol{footnote}}
 \def\@makefnmark{\hbox
 to 0pt{$^{\@thefnmark}$\hss}}
 \if@twocolumn
 \twocolumn[\@maketitle]
 \else \newpage
 \global\@topnum\z@ \@maketitle \fi\thispagestyle{plain}\@thanks
 \endgroup
 \setcounter{footnote}{0}
 \let\maketitle\relax
 \let\@maketitle\relax
 \gdef\@thanks{}\gdef\@author{}\gdef\@title{}
 \gdef\@affiliation{} \let\affiliation\relax	%
 \let\thanks\relax}

\def\@maketitle{\newpage
 \null
 \vskip 0em plus 2em minus 0em     % Vertical space before date.
 \ifx\@date\@empty\else
   \begin{flushright}
    {\ifnoncomplete(\today)
     \else{{\normalsize \@date}\\}\fi}      % \today printed
   \end{flushright}
   \vskip 3em plus 2em minus 2em   % Vertical space above title.
 \fi
 \begin{center}
  {\frtnsfb \@title \par}     % Title set in \Large size of \sf.
  \vskip 3em plus 1em minus 1.5em  % Vertical space after title.
  {%\large                        % each author set in \large, in a
   \lineskip .5em plus 0em minus .3em   % tabular environment
   \begin{tabular}[t]{c}\@author
   \end{tabular}\par}
\end{center}
 \par
 \vskip 6em plus 2em minus 4em}     % Vertical space after author.

\def\abstract{\if@twocolumn
\section*{Abstract}
\else \normalsize
\fi}

\def\endabstract{\if@twocolumn\fi\par\clearpage}

\def\section{\@startsection {section}{1}{\z@}{3.5ex plus 1ex minus
    .2ex}{2.3ex plus .2ex}{\normalsize\bf}}
\def\subsection#1{\subsectioncom{\sc{#1}}}
\def\subsectioncom{\@startsection{subsection}{2}{\z@}
    {3.25ex plus 1ex minus .2ex}{1.5ex plus .2ex}{\small}}
\def\subsubsection{\@startsection{subsubsection}{3}{\z@}{3.25ex plus
1ex minus .2ex}{1.5ex plus .2ex}{\small}}

\def\@addmarginpar{\@next\@marbox\@currlist{\@cons\@freelist\@marbox
    \@cons\@freelist\@currbox}\@latexbug\@tempcnta\@ne
    \if@twocolumn
        \if@firstcolumn \@tempcnta\m@ne \fi
    \else
      \if@mparswitch
         \ifodd\c@page \else\@tempcnta\m@ne \fi
      \fi
      \if@reversemargin \@tempcnta -\@tempcnta \fi
    \fi
    \ifnum\@tempcnta <\z@  \global\setbox\@marbox\box\@currbox \fi
    \@tempdima\@mparbottom \advance\@tempdima -\@pageht
       \advance\@tempdima\ht\@marbox \ifdim\@tempdima >\z@
      \else\@tempdima\z@ \fi
    \global\@mparbottom\@pageht \global\advance\@mparbottom\@tempdima
       \global\advance\@mparbottom\dp\@marbox
       \global\advance\@mparbottom\marginparpush
    \advance\@tempdima -\ht\@marbox
    \global\ht\@marbox\z@ \global\dp\@marbox\z@
    \vskip -\@pagedp \vskip\@tempdima\nointerlineskip
    \hbox to\columnwidth
      {\ifnum \@tempcnta >\z@
          \hskip\columnwidth \hskip\marginparsep
        \else \hskip -\marginparsep \hskip -\marginparwidth \fi
       \box\@marbox \hss}
    \vskip -\@tempdima
    \nointerlineskip
    \hbox{\vrule \@height\z@ \@width\z@ \@depth\@pagedp}}

\def\ref#1{%{\ifnoncomplete\marginpar{\footnotesize\sl ref [#1]}\fi}%
    \@ifundefined{r@#1}{{#1}\@warning{Reference `#1'
    on page \thepage \space
    undefined}}{\edef\@tempa{\@nameuse{r@#1}}\expandafter
    \@car\@tempa \@nil\null}}
\def\refn#1{\@ifundefined{r@#1}{{#1}\@warning{Reference `#1'
    on page \thepage \space
    undefined}}{\edef\@tempa{\@nameuse{r@#1}}\expandafter
    \@car\@tempa \@nil\null}}

\def\endequationl{\eqno \@eqnnum % $$ BRACE MATCHING HACK
$$\global\@ignoretrue}

\def\eqnarray{\stepcounter{equation}\let\@currentlabel=\theequation
\global\@eqnswtrue
\global\@eqcnt\z@\tabskip\@centering\let\\=\@eqncr
$$\arraycolsep\z@
\halign to \displaywidth\bgroup\@eqnsel\hskip\@centering
  $\displaystyle\tabskip\z@{##}$&\global\@eqcnt\@ne
  \hskip 2\arraycolsep \hfil$\displaystyle{{}##{}}$\hfil
  &\global\@eqcnt\tw@ \hskip 2\arraycolsep
  $\displaystyle\tabskip\z@{##}$\hfil
   \tabskip\@centering&\llap{##}\tabskip\z@\cr}
\def\mmodetrue{\mmode=\iftrue}

\def\eqnarrayl#1{\stepcounter{equation}\let\@currentlabel=\theequation
\label {#1}
\global\@eqnswtrue
\global\@eqcnt\z@\tabskip\@centering\let\\=\@eqncr
$$\arraycolsep\z@
\halign to \displaywidth\bgroup\@eqnsel\hskip\@centering
  $\displaystyle\tabskip\z@{##}$&\global\@eqcnt\@ne
  \hskip 2\arraycolsep \hfil$\displaystyle{{}##{}}$\hfil
  &\global\@eqcnt\tw@ \hskip 2\arraycolsep
  $\displaystyle\tabskip\z@{##}$\hfil
   \tabskip\@centering&\llap{##}\tabskip\z@\cr}
\def\label#1{
\@bsphack\if@filesw {
{\ifnoncomplete{\makebox[1in][r]{\footnotesize{\sl [#1]}}}\fi}%
\let\thepage\relax
   \xdef\@gtempa{\write\@auxout{\string
      \newlabel{#1}{{\@currentlabel}{\thepage}}}}
}\@gtempa
   \if@nobreak \ifvmode\nobreak\fi\fi\fi\@esphack}

\def\newlabel#1#2{
\@ifundefined{r@#1}{}{\@warning{Label `#1' multiply
   defined}}\global\@namedef{r@#1}{#2}}

\def\endeqnarrayl{\@@eqncr\egroup
      \global\advance\c@equation\m@ne$$\global\@ignoretrue}

\newif\if@numbersec \@numbersectrue
\def\appendix{\par\clearpage
  \setcounter{section}{0}
  \setcounter{subsection}{0}
  \def\thesection{\Alph{section}}
  \def\thesubsection{\arabic{subsection}}
  \@ifstar{\def\@sectname{Appendix}\@numbersecfalse}
          {\def\@sectname{Appendix~}\@numbersectrue}}

\def\thefigures#1{\par\clearpage\section*{Figures\@mkboth
  {FIGURES}{FIGURES}}\list
  {Fig.~\arabic{enumi}.}{\labelwidth\parindent\advance\labelwidth -\labelsep
      \leftmargin\parindent\usecounter{enumi}}}

\def\thetables#1{\par\clearpage\section*{Tables\@mkboth
  {TABLES}{TABLES}}\list
  {Table~\arabic{enumi}.}{\labelwidth-\labelsep
      \leftmargin0pt\usecounter{enumi}}}

\def\@sect#1#2#3#4#5#6[#7]#8{\ifnum #2>\c@secnumdepth
     \def\@svsec{}\else
     \refstepcounter{#1}\edef\@svsec{\ifnum #2=1 \@sectname
         \if@numbersec\csname the#1\endcsname\fi.\else
         \csname the#1\endcsname.\fi
        \hskip 1em }\fi
     \@tempskipa #5\relax
      \ifdim \@tempskipa>\z@
        \begingroup #6\relax
          \@hangfrom{\hskip #3\relax\@svsec}{\interlinepenalty \@M #8\par}
        \endgroup
       \csname #1mark\endcsname{#7}\addcontentsline
         {toc}{#1}{\ifnum #2>\c@secnumdepth \else
                      \protect\numberline{\csname the#1\endcsname}\fi
                    #7}\else
        \def\@svsechd{#6\hskip #3\@svsec #8\csname #1mark\endcsname
                      {#7}\addcontentsline
                           {toc}{#1}{\ifnum #2>\c@secnumdepth \else
                             \protect\numberline{\csname the#1\endcsname}\fi
                       #7}}\fi
     \@xsect{#5}}

\def\@sectname{}

 \def\thefootnote{\fnsymbol{footnote}}

\def \@magscale#1{ scaled \magstep #1}
\font\frtnsfb = cmssbx10 \@magscale2 % bold sans serif
\newcount\ieq
\newcount\jeq
\newcount\keq
\def \eq{
\multiply\ieq by 2
\jeq=\ieq
\divide\jeq by 4
\multiply\jeq by 4
\ifnum\ieq=\jeq \end{eqnarray} \keq=1 %\typeout{ Ieq.eq.jeq true.}
\else
\keq=2 \begin{eqnarray} \fi
\ieq=\keq
}

\def \mathbox(#1){\invisible\ifmmode{{#1}}\else{\mbox{${#1}$}}\fi}
\def \mbf(#1){\mbox{\boldmath{$#1$}}}
\def \abs(#1){\mathbox(\left|{#1}\right|)}
\def \bracket(#1){\mathbox(\left\langle{#1}\right\rangle)}
\def \brav(#1){\mathbox(\langle {#1}|)}
\def \cg(#1,#2,#3,#4,#5,#6){\mathbox({(#1\,#2\,#3\,#4|#5\,#6)})}
\def \comm(#1,#2){\mathbox(\left[{#1},{#2}\right])}
\def \dfdx(#1,#2){\mathbox(\frac{{\rm d}{#1}}{{\rm d}{#2}})}
\def \delfdelx(#1,#2){\mathbox(\frac{\partial{#1}}{\partial{#2}})}
\def \inprod(#1,#2){\mathbox({(#1\cdot #2)})}
\def \inprodij(#1){\mathbox({\inprod(#1_i,#1_j)})}
\def \intd(#1,#2){\mathbox({\int^#1_#2 \; \rmd})}
\def \eps(#1){\mathbox(\epsilon_{#1})}
\def \half(#1){\mathbox(\frac{#1}{2})}
\def\onehalf{\half(1)}
\def \ketv(#1){\mathbox(|{#1}\rangle)}
\def \matele(#1,#2,#3){\mathbox(\left\langle {#1}|\,{#2}\,|{#3}\right\rangle)}
\def \mateled(#1,#2,#3){\mathbox(\left\langle
{#1}||\,{#2}\,||{#3}\right\rangle)}
\def \hatmbf(#1){\mathbox({\hat{\mbf({#1})}})}
\def \ninej(#1,#2,#3,#4,#5,#6,#7,#8,#9){\mathbox(\left\{\matrix
     {#1&#2&#3\cr#4&#5&#6\cr#7&#8&#9\cr}\right\})}
\def \rtov(#1,#2){\mathbox(\sqrt{{#1\over #2}})}
\def \sixj(#1,#2,#3,#4,#5,#6){\mathbox(\left\{\matrix
     {#1&#2&#3\cr#4&#5&#6\cr}\right\})}
\def \third(#1){\mathbox(\frac{#1}{3})}
\def \Trace(#1){\mathbox({\hbox{Tr} \left\{#1\right\}})}
\def \outprod(#1,#2){\mathbox({(#1\times #2)})}

\def \als{\mathbox(\alpha_s)}
\def \bra{\mathbox(\langle)}

\def \etal{{\it et al.}}
\def \etc{{\it etc.}}

\def \heart{\mbox{$\heartsuit$}}
\def \ie{{\it i.e.}}
\def \invisible{\mbox{$\rule{0mm}{1mm}$}}
\def \ket{\mbox{$\rangle$}}
\def \lamlam{\mbox{$(\lam_i\cdot\lam_j)$}}
\def \LamQCD{\mbox{$\Lambda_{\rm QCD}$}}
\def \Lam{\mbox{$\Lambda$}}
\def \lam{\mbox{$\lambda$}}
\def \Nc{\mbox{$N_c$}}
\def \Nf{\mbox{$N_f$}}

\def \psibar{\mathbox(\bar{\psi})}
\def \qbar{\mbox{$\bar q$}}
\def \rmd{{\rm d}}

\def \Sig{\mbox{$\Sigma$}}
\def \sig{\mbox{$\sigma$}}
\def \simg{\raisebox{0.3ex}{$>$}\hspace{-0.8em}\raisebox{-0.3em}{$\sim$}}
\def \siml{\raisebox{0.3ex}{$<$}\hspace{-0.8em}\raisebox{-0.3em}{$\sim$}}

\def \spade{\mbox{$\spadesuit$}}
\def \TS{\mbox{$\rule[-3mm]{0mm}{8mm}$}}

\def \vecr{\mathbox({\vec r})}

\makeatother

\begin{document}

%%  local definitions
\def\Ellh{\Heff^{(3)}}
\def\Ellt{\Heff^{(2)}}
\def\Ell{{\cal L}_{\rm eff}}
\def\Heff{H}
\def\Nc{N_c}
\def\Nf{N_f}
\def\OGE{{\rm OGE}}
\def\Vinsh{V^{(3)}_{\rm III}}
\def\Vinst{V^{(2)}_{\rm III}}
\def\Voge{V_{\rm CMI}}
\def\piii{p_{\rm III}}
\def\als{\alpha_s}
\def\etal{{\it et al.}}
\def\etc{{\it etc.}}
\def\ie{{\it i.e.}}
\def\lamlam{\lam_i\cdot\lam_j}
\def\lam{\lambda}
\def\meff{m^{\rm eff}}
\def\phid{\phi^{\dagger}}
\def\psiL {q_L}
\def\psibR{\qbar_R}
\def\psib{\qbar}  \def\qb{\qbar}
\def\psizb{\bar\psiz}
\def\psiz{q_{(0)}}
\def\qbar{\bar q}
\def\qqbar{q\bar q}
\def\rc{\rho_c}
\def\rv{\vec r}
\def\sigsig{\sigv_i\cdot\sigv_j}
\def\sigv{\vec\sigma}
\def\sig{\sigma}
\def\tauv{\vec\tau}
\newcommand{\KY}{K.~Yazaki}
\newcommand{\MO}{M.~Oka}
\renewcommand{\TS}{\mbox{$\rule[-3mm]{0mm}{9mm}$}}
\renewcommand{\bra}{\langle}
\renewcommand{\ket}{\rangle}
\renewcommand{\psibar}{\bar{\psi}}
\newcommand{\rhat}{\hat{r}}
\renewcommand{\rtov}[2]{\mbox{$\sqrt{\frac{#1}{#2}}$}}
\renewcommand{\simg}{\raisebox{0.3ex}{$>$}
                   \hspace{-0.8em}\raisebox{-0.3em}{$\sim$}}
\renewcommand{\siml}{\raisebox{0.3ex}{$<$}
                   \hspace{-0.8em}\raisebox{-0.3em}{$\sim$}}
\renewcommand{\spade}{\mbox{$\spadesuit$}}
\renewcommand{\third}{\mbox{$\frac{1}{3}$}}
\def\Gam{{\mit \Gamma}}
\def\III{\mbox{\rm I\thinspace I\thinspace I}}
\def\IIId{\mbox{\rm I\thinspace I\thinspace I$_2$}}
\def\IIIt{\mbox{\rm I\thinspace I\thinspace I\hspace{0.2mm}3}}
\def\Omegad{\mathbox({\mit \Omega}_2)}
\def\Omegat{\mathbox({\mit \Omega}_3)}
\def\VIII{\mathbox(V_{\rm III})}
\def\VIIId{\mathbox(V_{\rm III2})}
\def\VIIIt{\mathbox(V_{\rm III3})}
\def\Voge{\mathbox({V_{\rm OGE}})}
\def\Vcmi{\mathbox(V_{\rm CMI})}
\def\Vconf{\mathbox(V_{\rm CONF})}

\def\bigbra{\left\langle}
\def\bigket{\right\rangle}
\def\xiij{\xi_i\xi_j}
\def\pro{\protect}
\def\wave{\simeq}
\def\Rv{{\vec R}}

%\nonfinal \def\heart{\spade}
\final \def\heart{}

\date{TIT/HEP-283/NP\\
      hep-ph/9502295\\
      February, 1995}

\title{Strong and Weak Interactions of Strange Hadrons%
\footnote{talk presented  at {\sl the YITP Workshop on
``{\it From Hadronic Matter to Quark Matter}''}, Kyoto, October 30 --
November 1, 1994; to be published in Prog.\ Theor.\ Phys.\ Suppl.}}

\author{%
Makoto Oka\thanks{e-mail: oka@th.phys.titech.ac.jp}\\
{\sl Department of Physics, Tokyo Institute of Technology}\\
{\sl Meguro, Tokyo 152, Japan}\\
}

\maketitle

\abstract {
I review  hadronic processes involving  strange hadrons,
especially hyperons from the quark structure point of view.  The
strong interaction of quarks expects several important new features
when the strangeness is introduced upon the non-strange
degrees of freedom.   I concentrate on the instanton induced
interaction, effects of the Pauli principle for strangeness and the
nonleptonic weak interaction of strangeness.
The single hadron as well as the two baryon systems including the $H$
dibaryon and other exotic hadronic systems with strangeness are
studied in this context.}

\thispagestyle{empty}

\newpage
\def\SU#1{$SU(#1)$}
\def\SULR#1{$SU(#1)_L\times SU(#1)_R$}
\def\UA{$U_A(1)$}

\section{Introduction}

Strangeness is the third lightest flavor whose mass is comparable to
the QCD scale parameter, \LamQCD.  The chiral \SULR3\  seems powerful in
studying the low energy hadron phenomena, although
the large strange quark mass breaks the \SU3\ symmetry significantly.
Because the QCD coupling strength depends on the energy scale, QCD is
not quite flavor blind.  In fact, the light quark dynamics is very
different from that of  heavy quarks.  The heavy quark symmetry
reveals that the light degrees of freedom can be separated from the
heavy ones in many of  heavy hadron phenomena.

Is the strange quark dynamics just a repetition of the light one with
 complication due to the symmetry breaking?
I here argue that some new important and
interesting physics issues arise from the introduction of the new
flavor.
I itemize them and will explain (most of) them later.

(a) The \UA\ anomaly rpresented by the instanton-induced interaction,
which enhances the flavor mixing in the pseudoscalar mesons
and plays significant roles in multihadron systems.

(b) The Pauli ``allowance'' in the multi-hadron systems or the matter,
which makes the strange quark matter, the $H$ dibaryon, heavy multistrange
nuclei plausible.

(c) The strangeness changing weak processes, which are
experimentally studied in  hypernuclear decays and reveal new features
of  hadronic weak interactions.

Strangeness plays major roles in various stages of the hadron physics.

(1) First, the single hadron spectrum is (historically) important in
revealing the \SU6\ symmetry breaking and
the chiral \SULR3\  symmetry.  There the symmetry breaking
is minimal as is seen for instance in the baryon magnetic
moments. Namely, the \SU3\ or \SU6\ symmetry is good for  the currents
and wave functions for mesons and baryons.  This makes the naive
valence quark model successful.  At the same time  the
explicit symmetry breaking and the \UA\ anomaly are found to
play important roles in the
single hadron system.

(2) When we turn to hadronic interactions, a new aspect of the third
flavor becomes important.  It is the Pauli exclusion principle.
Strangeness in  nonstrange environment is much freer than the
nonstrange quarks and thus the spectrum is richer than the nonstrange
systems.  One such example is the $H$ dibaryon, which is a strangeness
$-2$ six-quark bound state.  I will also show that the hyperon-nucleon
interactions  exhibit a variety of  interesting features of hafronic
interactions.
Experimental efforts on the hyperon and hypernuclear physics have
developed a
new exciting field in hadron physics.

(3) In hadronic matter, various interesting possibilities have been suggested,
such as, the strange quark matter, multistrange heavy nuclei, kaon
condensation, K balls.  Here again the symmetry breaking is
important.  Comparing to the pion, the kaon is relatively heavy.  This
makes some of the strange systems more stable than the corresponding
nonstrange systems.

(4) Weak interactions of strangeness reveal new roles of hadronic
interactions.  The $\Delta I= \onehalf$ rule, for instance, is
experimentally confirmed in various single hadron decays, and yet its
mechanism is not fully understood.
Intereference between th standard theory of the weak interaction and
the strong hadronic interaction is to be studied from this viewpoint.

In this article, I would like to review some of the above mentioned
``strange'' subjects in order.

\section{Strangeness in Hadron}

\subsection{instanton induced interaction}

The instanton is a gluon field
configuration with nontrivial topology and
has provided a qualitative understanding of various
nonperturbative features of QCD{\cite{instanton}}.
One of its important consequences is that the
coupling of light quarks ($u$, $d$ and $s$)
to the instanton breaks the axial \UA\
symmetry and causes the $\eta$-$\eta'$ mass splitting{\cite{tH}}.

Dynamical properties of instantons in the QCD vacuum have been
studied by Shuryak and his collaborators\cite{Sh}.
They found that the QCD vacuum can be  described by an instanton
liquid composed of small-size ($\wave 0.3$ fm) instantons and
anti-instantons.
The density of instantons is low ($\wave 1$ fm$^{-4}$) so that
orientations of near-by instantons are almost independent from each other.
This gives a gluon condensate consistent with the QCD sum rule and a
quark condensate as a result of  dynamical breaking of the \SULR3\
chiral symmetry.
The light quarks get constituent masses of a few hundred
MeV due to the symmetry breaking.

A recent lattice QCD calculation tested the instanton liquid picture
of the QCD ground state with the help of the cooling technique{\cite{Negele}}.
After a cooling it has been shown that only a few instantons and
anti-instantons survive and yet the meson and baryon correlators are
essentially unchanged from the original uncooled calculation.
It thus indicates that the instantons in the vacuum play the most
essential role in determining the low lying mesons and baryons.

The \UA\ symmetry breaking is attributed to the instanton-light
quark coupling\cite{tH,SVZ}.
't Hooft showed that a massless quark forms a zero mode around the instanton,
which dominates the light-quark instanton coupling at low energy.
In the instanton liquid vacuum, quarks form a zero mode around each
instanton independently and propagate by hopping from one zero mode to
another.

The effective interaction of light quarks in the instanton vacuum,
called instanton induced interaction \III,
is given by a local effective interaction\cite{tH,OT,OTH},
\eq
      \Ellh  = V_0 \,
           \psibR(1) \psibR(2) \psibR(3)  \, {189 \over 40}
           \,{\cal A}_3^f\,
              \bigl ( 1 - {1\over 7} \sum_{i>j=1}^3 \sigsig \bigr)
           \psiL(3) \psiL(2) \psiL(1)
            + \hbox{(h.c.)}
\label{eq:H3}
\eq
Here
$q_{L(R)} \equiv \half (1\mp\gamma_5)\, q$
is the left (right)  handed quark field operator, and
${\cal A}^f_n$
is the projection operator to the flavor singlet state
of $n$ quarks.
The hermitian conjugate term (h.c.) arises from
the interaction through the anti-instanton.
The strength $V_0$ depends on the density of the instanton in the
QCD vacuum.

The $\bar R$(right-handed) -- $L$(left-handed) structure of eq.(\ref{eq:H3})
clearly shows the chirality nonconservation.
It however applies only to the flavor singlet part because \III\ is
proportional to the projection ${\cal A}^f_3$ into the flavor singlet
states.
Thus the \UA\ symmetry is broken by \III, while the \SULR3\
is preserved.

The six-quark \III\ induces a four quark vertex \IIId\
through the contraction of a pair of $\psibR$ and $\psiL$ by the current
quark mass term or by the quark condensate in vacuum,
\eq
\Ellt =
       V_{0}^{(2)} \, \psibR(1) \psibR(2) \, {15 \over 8} {\cal A}_2^f
\, ( 1- {1 \over 5} \sigv_1\cdot\sigv_2 ) \,\psiL(2) \psiL(1) +
\hbox{(h.c.)}
\label{eq:H2}
\eq
The strength of the two-body interaction
$V_0^{(2)}$ is related to the three-body strength
and the quark condensate of the third flavor \cite{SVZ},
\begin{equation}
V_0^{(2)} (1,2) \equiv
           {1\over 2} V_0\,(\bra\bar q q\ket -K \tilde m_3)
           = {1\over 2} V_0\,(-K m_3)
\end{equation}
where $K$ is a positive constant and $\tilde m_i$ ($m_i$) is the
current (constituent) quark mass of the $i$-th flavor.
The \SU3\ symmetry breaking is represented here by the strange quark
mass so that the interaction between the $u-d$ quarks is stronger than
that for $s-u$ or $s-d$.
The ratio,
$V_0^{(2)}(u,s)/V_0^{(2)}(u,d)$,
is given by $\xi \equiv m_u/m_s$.
The hadron spectrum indicates $\xi\approx 0.6$, while the QCD sum rule
suggests $\xi\approx 0.8$\cite{Sh}.
Note that \IIId\ is attractive, while the full \III, eq.(\ref{eq:H3}),
is repulsive, because
$\bra\qbar q\ket$ is negative.
%We take $\bra\bar qq\ket = - $ (250 MeV)$^3$,
%which is obtained by the QCD sum rule.

\subsection{Meson}

Many studies have been carried out for the meson spectrum in the
three-flavor Nambu-Jona-Lasinio (NJL) model{\cite{NJL}}.
This model is particularly useful in understanding the pseudoscalar
mesons which are regarded as the Nambu-Goldstone (NG) bosons for the chiral
symmetry breaking.
There the 6-quark \III, eq.(\ref{eq:H3}), is commonly used to
represent the \UA\ anomaly.
For the $\qqbar$ meson system, the loop contraction of a $\qqbar$ pair
leads to the 4-quark effective vertex \IIId, eq.(\ref{eq:H2}).
As \IIId\ requires the projection to the flavor singlet (antisymmetric)
state, it contributes to the $\eta_1$ meson repulsively, while $\pi$
and $\eta_8$ get strong attraction.  Thus the \UA\ symmetry is broken
explicitly by \III.

The mixing of $\eta_1$ and $\eta_8$ comes mainly from the \SU3\
symmetry breaking.  The large strange quark mass makes ($s\bar s$)
separated from ($u\bar u$) and ($d\bar d$).  Without the \UA\
breaking, the ideal mixing, \ie, $\eta' = s\bar s$ and
$\eta = (u\bar u
+ d\bar d)/\sqrt{2}$ would be realized, while \IIId\ mixes the $s\bar s$
with $(u\bar u + d\bar d)/\sqrt{2}$.
Thus the instanton induced interaction and the \SU3\ breaking due to
the quark mass work in the opposite direction for the flavor mixing.

Due to the $\bar R$ -- $L$ structure, the \IIId,
has the
opposite sign for the scalar and pseudoscalar channels.  It turns out
that the large flavor mixing in the $\eta-\eta'$ system is consistent
with the instanton induced interaction.
Namely, the \IIId\ is strongly attractive in the pseudoscalar channel
so that the chiral symmetry is dynamically broken and generates
the $\bra\qbar q\ket$ condensate, and the  pseudoscalar octet mesons
$\wave (\qbar (\lam_i/2)\gamma^5 q)$ are transmuted into the massless
(light) NG bosons.
In the scalar channel
the flavor mixing is suppressed and therefore the OZI separation
is achieved.  The mixings in the vector and the axial vector channels
are also suppressed to order 1/$\Nc$.
Thus, the OZI rule is well satisfied in the scalar, vector and
axialvector channels,  while the pseudoscalar mesons have a large
flavor mixing.
In conclusion, the light-quark-instanton interaction provides a
consistent picture of the \UA\ breaking and the flavor mixing in
various channels.

\subsection{Baryon}

The baryon mass spectrum and the magnetic moments support the
spin-flavor \SU6\
symmetry.   Also the meson-baryon couplings are consistent with the
\SU3\ (and \SU6) symmetry of the wave functions.
The explicit breaking of \SU3\ is minimal and is taken care by the
Gell-Mann-Okubo formalism, which treats the \SU3\ octet term in the
hamiltonian perturbatively.

Contrary to the mesons, which are deeply bound and thus are to be
treated relativistically,
the conventional constituent quark model gives a simple and good
picture for baryons.
Three valence constituent quarks confined in a potential well interact
with each other by exchanging gluons.
The one gluon exchange interaction contains a color-magnetic term,
that breaks the \SU6\ symmetry.
A relevant term of OGE in the following discussion is
the central part of the color magnetic interaction (CMI),
which reads in the Breit-Fermi form
\begin{equation}
   \Vcmi = - {\als\over 4} \sum_{i<j}
              {2\pi\over 3m_im_j} \, (\lamlam)  \, (\sigsig)\,
                  \delta(\rv_{ij})
\label{eq:CMI}
\end{equation}
The $N-\Delta$  and the $\Lam-\Sig$ mass difference both arise from
this term, providing that the ratio of the strange and nonstrange
quark masses are about 0.6.  The same interaction deforms the nucleon
wave function, which results in the nonvanishing neutron charge radius.

What are the roles of the instanton induced interaction in the baryon?
For the valence quark model, we reduce  \III, eqs.(\ref{eq:H3}) and
(\ref{eq:H2}), into  the following nonrelativistic forms:
\begin{eqnarray}
      \VIIIt &=&  V_0 \,
           {189 \over 40}
	   \sum_{(ijk)}\, {\cal A}^f_3\,
              \left[ 1 - {1\over 7}
(\sigma_i\cdot\sigma_j+\sigma_i\cdot\sigma_k+\sigma_j\cdot\sigma_k) \right]
\delta(\vecr_{ij}) \delta(\vecr_{jk})
\label{eq:V3}
\\
   \VIIId    &=& U_0^{(2)}\, \sum_{i<j}\, {1\over m_im_j}
          \left[ 1 +
        {3\over 32} \lam_i\cdot\lam_j + {9\over 32}
           \sigv_i\cdot\sigv_j \, \lam_i\cdot\lam_j \right]
                          \delta(\vecr_{ij})
\label{eq:V2}
\end{eqnarray}
where $ U_0^{(2)}=-\onehalf V_0K m_u^2m_s$,
and $m_i$ is the constituent quark mass.

It is interesting to note that the two-body instanton-induced
interaction contains the identical term to eq.(\ref{eq:CMI}), and
thus is not distinguishable from the one-gluon exchange as far as the
hyperfine splittings are concerned.
We recently proposed a view that both the instanton induced
interaction and the one-gluon exchange interaction share the hyperfine
splittings of a meson-baryon spectrum{\cite{OT,OTH,sac}}.
For simplicity, we fix the total strength of the spin-spin interaction
so as to reproduce the nucleon delta mass difference, which is the
typical hyperfine splitting in the baryon spectrum.
We represent the share of
the instanton induced interaction by $\piii$, i.e.,
\eq   H_{hyp} = \piii \VIIId +  (1-\piii) \Vcmi .
\eq
In order to determine $\piii$, the $\eta-\eta'$ mass difference is
calculated.  We find that $\piii \approx 30 - 40$\% gives a reasonable
mass difference\cite{OT}.

It is also important to notice that the three-body instanton-induced
interaction has no direct contribution to the  valence quarks in the baryon,
if the small size instanton is assumed.
This is the result of the symmetry consideration. The three valence
quarks are color antisymmetric and must be flavor antisymmetric,
orbital symmetric for the instanton induced interaction.  But this
combination is not allowed for the fermion.

This leads to the conclusion that contrary to the mesonic case
the instanton induced interaction
does not disturb the single baryon spectrum successfully explained in
the conventional quark models.

\section{Strangeness in $B=2$}

When strangeness is planted into  hadronic matter, a new important
effect comes into effect.  It is the Pauli principle.
Witten first suggested that the strange quark matter (strangelet) can
be the most stable system with finite baryon number, because the Pauli
principle prefers  mixing of strangeness in the ordinary
matter{\cite{Witten}}.
The ratio of the strangeness and the baryon number
is of order one in the strangelet.
Despite the heavier strange-quark mass, large
strangeness mixing is favored because of  the high Fermi energy for the
ordinary quarks in bulk matter.

The quark antisymmetrization plays a major role even in $B=2$ systems
with strange\-ness.  In fact, the reason why the deuteron binding energy
is much smaller than the energy scale of ordinary hadronic interaction
can be understood by the symmetry consideration.
It is the strong short-range repulsion that prevents the deuteron, or
any other nuclei, from crashing into quark matter.
The simple quark model symmetry predicts that the Pauli principle with
the help of the hyperfine interaction among quarks yields strong
short-range repulsion in most of two-baryon systems{\cite{Nagoya}}.
Introduction of new flavor, the strangeness, may reduce the Pauli
exclusion requirement, \ie, the Pauli allowance and thus makes 6-quark
type dibaryons possible to exist.  The most plausible candidate is the
$H$ dibaryon.

\subsection{$H$ dibaryon}

The $H$ dibaryon is a bound state of six quarks, or two baryons with
strangeness $-2$.
The strong hyperfine
interaction due to the gluon exchange favors the flavor singlet
6-quark state{\cite{Jaffe}}, and indeed
several quark model calculations predict a bound state, $H$, with
$B=2$, $J^{\pi}=0^+$, and strangeness $-2$ \cite{QCMH,Hdibaryon}.
$H$ couples to the
ordinary two baryon channels, $\Lam\Lam$, $N\Xi$, and $\Sig\Sig$.
If the $H$ mass is below the $\Lam\Lam$ threshold,  $H$ decays (into $N
\Sig$) only by the weak interaction.

Experimental searches  of $H$ are underway in two facilities, BNL and
KEK.  The KEK group has found  two candidates for
double hypernuclei, $_{\Lambda\Lambda}A$, which are formed in the
$(K^+,K^-)$ experiment on the emulsion target.
The estimated binding energy of the double hypernuclei places the
upper limit of the $H$ binding energy below the $\Lam\Lam$ threshold.
Thus,  the
possibility of deeply bound $H$ with binding energy 30 MeV or more
has been excluded.
Existence of $H$ should have a significant effect in the formation of  double
hypernuclei and $\Xi$-hypernuclei.
Sakai \etal\ studied the interaction between $H$ and
nuclei in the quark model and suggested possibility of $H$ bound
weakly to
nuclear matter{\cite{Sakai}}.

When the binding energy of $H$ is small, we cannot neglect the effects of two
baryon continuum states.  The quark cluster model (QCM) calculation is
more appropriate for this situation than the bag model,
as one can couple the $\Lam\Lam$, $N\Xi$, and $\Sig\Sig$ channels with $H$.
The QCM calculation for $H$ has been carried out by several
groups{\cite{QCMH,Hdibaryon}}.
The results, which seem sensitive to the choice of the long-range meson
exchange interaction, indicate that $H$ is stable
with the binding energy ranging from 40 to 120 MeV unless the
instanton effect is taken into account.
We, however, found that $H$ as the six-quark system with strangeness has
significant contribution from the 3-body  instanton induced
interaction and that \III\ in this channel is strongly
repulsive{\cite{OTH}}.
The effect is more than 50 MeV, and
results in an unstable or barely bound $H$ dibaryon.
It has also been  found that the flavor non-singlet component $H'$ (
with the flavor [42] symmetry) becomes almost degenerate with the flavor
singlet $H$ for a strong instanton induced interaction{\cite{sacH}}.

We have to wait for further experimental data to make a conclusion,
but it is plausible now that $H$ is not a deeply bound 6-quark state, but
a weakly bound state of $\Lam\Lam$ and $N\Xi$.
In fact, the flavor singlet combination is dominated by $N\Xi$ and the
quark cluster model analysis supports the large $N\Xi$ contribution in
$H$.

\subsection {Hyperon-Nucleon Interaction}
Study of hypernuclei has revealed that the hyperon bound in nuclear
matter has a binding energy less than that of the nucleon.  This fact
clearly suggests that the hyperon-nucleon (YN) interaction is as strongly
repulsive as the nuclear force at short distances, because the
long-range part of the interaction should not be much weaker for YN.
Low energy YN scattering data support the existence of the
short range repulsion between YN.

The phenomenological YN potentials{\cite{Nij,Julich}} based on the
\SU3-symmetric meson
exchange interaction usually assume a repulsive core or a form factor
cutoff  to represent the repulsion.  The core radii or the  cutoff
masses are free parameters to be determined phenomenologically in each
channel.   Because the number of experimental data is so
small that the fit is not always unique.  Different potential models
have sometimes very different core radii.

\def\wave{\simeq}
\def\rtt{\sqrt{3}}
\def\Rv{{\vec R}}
\def\MO{M.~Oka} \def\KY{K.~Yazaki}

On the other hand, the quark model description of the short-range
repulsion between two nucleons is quite successful{\cite{OY80,QCM,Nagoya}}.
It has been demonstrated in the quark
cluster model (QCM) calculations that the color-magnetic gluon
exchange and the
quark antisymmetrization in the valence quark dynamics provides a
non-local soft
repulsive core, which can reproduce the $N-N$ scattering S matrices for
energies up to 300--400 MeV.
When QCM is applied to other two-baryon systems, the same
mechanism yields strong short-range
repulsions in most of the two octet-baryon systems, including \Lam-N
and \Sig-N{\cite{QCMYN,Nagoya}}.
We find that the quark
exchange effects (due to the antisymmetrization) show distinctive
spin-isospin dependences especially for the \Sig-N interactions.
Such strong channel dependences are attributed for the \SU6\ quark
symmetry of the two-baryon system and
have not been considered or taken into
account in the conventional YN potential models.
Thus it is quite interesting to see whether a realistic YN interaction
that incorporates the quark-exchange mechanism can reproduce the YN
two-body data as well as the hypernuclear spectrum.

Along this line of thought we
construct a QCM-based
potential model for the hyperon-nucleon interaction incorporating both
the meson exchanges and the quark-gluon effects{\cite{OSO}}.   Such a model
enables
us to analyze experimental YN scattering data and to determine whether
the quark-exchange mechanism is indeed at work for the hyperon-nucleon
systems.  This is important further for the study of double strange
systems, such as \Lam-\Lam, N-$\Xi$ and the H dibaryon, because the
interactions of $S=-2$ two-baryon systems are not yet directly accessible in
experiment and thus require theoretical predictions.

\subsection{Realistic YN Potential Model\protect\footnotemark}
\footnotetext{This section is
based on the research carried out by K.~Ogawa, S.~Takeuchi and
M.~Oka.\cite{OSO}}
We consider a valence quark model with a hamiltonian,
\eq    H=K+\Vconf +\Voge
\eq
where $K$ is the nonrelativistic quark kinetic energy term, $\Vconf$
stands for a quark confinement potential and $\Voge$ is the
Fermi-Breit potential for the one gluon exchange.
We employ the resonating group method (RGM) wave function for the
six-quark system, given by
\eq
    \Phi_{BB'} (1\sim 6) = {\cal A} [ \phi_B(1\sim 3)\,
\phi_{B'}(4\sim6) \,\chi(R)\,]
\eq
and solve the RGM integral equation, with the kernels $H$
(Hamiltonian) and $N$ (Normalization):
\eq
  \int \left[ H(R,R') - E\, N(R,R') \right] \, \chi(R')\, dR' = 0
\label{eq:QCM}
\eq
Nonlocality of the RGM equation comes from the antisymmetrization
of the quarks{\cite{OY80}}.

In order to describe the long-range part of the baryon-baryon
interaction, we additionally need the meson exchange potentials.
We keep the SU(3) symmetry for the meson-baryon couplings.  Indeed,
the $YN$ potential models, such as the Nijmegen models{\cite{Nij}} and
J\"ulich models{\cite{Julich}}, are based on the SU(3) symmetry.  In
this study, we employ a one-boson exchange potential based on
the Nijmegen potential model and instead of
using the hard cores in the original model, superpose it with the
quark exchange interaction at the short distance.
We introduce to the QCM equation
(\ref{eq:QCM})
the meson exchange potential,
which is of the form of the Nijmegen model.
This can be done by adding
an integral kernel for the meson exchange potential{\cite{QCM}}, given
by
\eq
       V(R,R') \equiv \int dR'' N^{1/2}(R,R'') V_{f} (R'')
      N^{1/2}(R'',R')
\eq
where $V_{f}$ is the meson exchange potential with the
appropriate form factor.  The form factor is chosen so as to be consistent
with the quark wave function of the baryon,
\eq
      V_{f} (R) \equiv \int \rho(x;R/2) V_{\rm meson} (x-y)
\rho(y;-R/2)\, dx\,dy
\eq
where $V_{\rm meson}$ contains the exchange potentials of the
pseudoscalar nonet, the vector nonet and the scalar mesons,
whose coupling constants satisfy the $SU(3)$ relations.
The form factor $\rho(x;R/2)$ is taken as
the quark density of the baryon centered at $R/2$, normalized as
\eq
	\int \rho(x)\,dx =1
\eq
In the QCM calculation, we employ the Gaussian for the
internal quark wave functions of the baryon for simplicity, and thus
the corresponding form factor is given also by a Gaussian.

Our main strategy is (1) to fix the model parameters, such as the
quark model parameters and the meson-baryon coupling constants, so as
to reproduce the $N-N$ scattering data and
(2) generate the $YN$ interaction using the \SU3\ symmetry and its
breaking due to the quark mass difference.
We then (3) compare the $YN$ scattering S matrices in our calculation
with those from the conventional potential models, such as the
Nijmegen models.

We find that the Pauli exclusion principle gives a stronger repulsion
for the $N\Sig$ ($S=0$, $I={1\over2}$) and
$N\Sig$ ($S=1$, $I={3\over2}$) channels,
while the other channels show a mild repulsion which is generally
softer than the original Nijmegen model D.
The repulsion in the $N\Sig$ ($I={3\over 2}$, $^3S_1$) channel is as
strong as that in the Nijmegen model F, which is known to provide not enough
binding for $\Sig$ to make a bound $\Sig$ hypernuclei.
The strong spin-isospin dependence of the $N-\Sigma$ interaction is
expected from the symmetry consideration in the quark cluster
model{\cite{Nagoya}}.
The stronger repulsions for ($S=0$, $I={1\over2}$) and
($S=1$, $I={3\over2}$) are attributed to the Pauli exclusion of the
six-quark state made of the completely overlapping $N-\Sig$ with these
quantum numbers.

We also calculate the YN cross sections and the polarizations in
order to study effects of the quark exchange repulsion to the
observables.
The low energy $\Lam-N$, and $\Sig-N$ scatterings have not been fully
studied experimentally.
A few available cross section data are compared with our results.
They suggest that the scalar meson exchange, which gives the major
part of the attraction, should contain both the flavor singlet and
octet components and their mixing.
Indeed, if we assume that the scalar $\sigma$ meson with the mass
about 550 MeV is flavor singlet, then the $\Lam-N$ and $\Sig-N$ cross
sections at low energy are too much enhanced.
Recent study of the $\Sig$ potential in our model by Takeuchi, Takayanagi and
Shimizu indicate that the QCM potential with the flavor singlet
$\sigma$ meson makes $^5_{\Sigma}He$ bound too much{\cite{Takayanagi}}.
Their conclusion is consistent with the hypernuclear spectroscopy, which
indicates that the mean field potential for $\Lam$ or $\Sig$ is weaker
than that for the nucleon.

Similar study of the $YN$ interaction has been carried out by the
T\"ubingen group{\cite{Tubingen}} and the
Niigata--Kyoto group{\cite{Fujiwara}}.
The T\"ubingen group proposed a QCM approach with the chiral scalar
($\sigma$) and the flavor-octet pseudoscalar mesons coupled directly
to valence quarks.  The $NN$ scattering phase shifts are fitted
successfully only after the $\sigma NN$ coupling constants are
modified for the $L\ge 1$ partial waves.  They also introduce baryon mass
dependences in the scalar-meson-exchange potential, which effectively
reduces the medium-range attraction in the $YN$ interactions.

The Niigata--Kyoto group employs the quark cluster model with a
scalar-nonet meson exchange
and the tensor part of the $\pi$ and $K$ exchanges.  They pointed out
that the flavor \SU3\ symmetry is preserved in  the global
feature of the interaction and argue that the attraction due to the
scalar meson exchange should be weaker for $\Lam N$ and $\Sig N$ so
that the effect of larger reduced mass is compensated.
Their potential model RGM--F based on the Nijmegen F potential can
successfully reproduce the $NN$ phase shifts as well as the available
$YN$ cross  section data{\cite{Fujiwara}}.

Dover and Feshbach studied
the \SU3\ symmetry structure of the $YN$
interactions{\cite{Feshbach}}.
They found that both the long-range meson exchange potentials and the
short-range quark-exchange repulsion contain the \SU3\ symmetry
breaking primarily confined to the diagonal transitions.  The
available $YN$ cross sections seem to support their conclusion.
This therefore indicates that the \SU3\ breakings in the meson and quark
exchanges come naturally from the meson/quark mass differences.

Compared with the progress made in these theoretical programs, the
experimental data for two-body YN interactions are yet poor.
While a recent measurement of the $\Sigma-N$ scattering cross section at KEK
is under analysis{\cite{Ieiri}}, further experimental study of the
two-body YN systems is desirable.

\subsection{Multistrange Matter}

The third flavor increases possibility of finding more exotic hadron
systems.
One of the examples is  multistrange hypernuclei.
Schaffner \etal{\cite{Multistrange}} predicted deeply bound
multistrange nuclei with the
strangeness fraction $\abs(S)/A \simeq 1$ and the binding energy per
baryon $E_B/A \ge 20 $ MeV.  Such systems contain many $\Xi$ baryons
because
the Pauli principle makes them stable against the strong decay,  $\Xi
N\to \Lam\Lam$.

Another strange hadron system of interest is
K-ball{\cite{Kball,Kball2}}, which is
a nontopological soliton (Q-ball) in the  chiral theory and is
considered as a (classical) bound state of many kaons.  The K ball with large
strangeness may be stable against strong decay modes and thus be a
long-lived mesonic object.
Possibility of such object was first studied by Distler
\etal{\cite{Kball}}, and
indeed the solutions are later constructed in a specific model, the
scale-invariant \SU3\ chiral effective theory{\cite{Kball2}}.
It was pointed out that the strangeness again favors such
nontopological soliton solutions, mainly because the kaon is much
heavier than the pion and therefore the kaon bound state may be more
stable than pions.
These exotic systems with multiple strangeness might be produced and
detected in high energy heavy ion collisions.  Indeed, it was
suggested{\cite{Dover-HI}} that coalescence of strange baryons in
heavy ion collisions has a significant probability of producing the
$H$ dibaryon if it is  a bound state.

\section{$\Delta S=1$ Weak Transition}

\def\wave{\simeq}
\def\rtt{\sqrt{3}}
\def\Rv{{\vec R}}
\def\MO{M.~Oka} \def\KY{K.~Yazaki}

Free $\Lambda$ decays weakly into a nucleon and a pion with
two isospin modes, $p\pi^-$ and $ n\pi^0$, which share 64\% and
36\% of the total decay rate.  If the decay goes through a $\Delta I=
{1\over 2}$ vertex, the $p\pi^-/n\pi^0$ ratio would be two to one except for
a small correction due to the phase space difference.
The experimental ratio is very close to the $\Delta I= {1\over 2}$
prediction, and thus support the $\Delta I= {1\over 2}$ hypothesis
for the hadronic weak decay.

In the nuclear medium, the $\Lambda\to N\pi$ decay is suppressed by
the Pauli blocking on the final nucleon state, whose momentum is
less than 100 MeV/c for the $\Lambda$ decay at rest.  Indeed, in heavy
hypernuclei, the decay is predominantly the nonmesonic one, that is,
$\Lambda N \to NN$.  If we assume that the initial $\Lambda$ and
$N$ are at rest, then the final relative momentum of $NN$ is
about 420 MeV/c and thus is well above the Fermi momentum.
Furthermore, this final momentum
is high enough to look into the
short-distance component of the two nucleon system
and therefore this type of the
hyperon decay may reveal a new aspect of the weak interaction of
quarks under
the influence of the strong interaction.
We here study the direct quark (DQ) processes in
the two-body $\Lambda N\to NN$ weak decays and
point out qualitative differences
from the conventional picture, \ie,  the meson ($\pi$, $K$, $\rho$,
etc.) exchange mechanism, where one
of the meson-baryon vertices involves the weak transition $s\to
d${\cite{meson}}.
It is quite timely to study this process in detail because new
accumulating experimental data have
revealed some difficulties in the meson-exchange picture.  For instance,
the so-called $n-p$ ratio, i.e., the ratio $R_{np}$ of $\Lambda n \to
nn$ v.s. $\Lambda p \to np$ decay in the nucleus, is predicted very
small,  $R_{np}\wave 0.1-0.4$ in the meson-exchange picture.  This is due
to the strong contribution of the tensor force, which is preferred at the
large momentum transfer.  The tensor force selects the $S=1$, $I=0$
$pn$ final state and therefore $R_{np}$ becomes small.
The experimental data seem not to agree with the prediction, i.e.,
$R^{exp}_{np} \wave 1$ in decays of light hypernuclei.
We argue that the direct quark process, which does not follow the
$I=0$ selection rule, may enhance the $n-p$ ratio.

Another important aspect of the nonmesonic weak decays is
the $\Delta I= \onehalf$ rule, which is known to be satisfied
in the mesonic weak decays of hyperons to about 5\%
error.   The same rule for the nonmesonic weak processes, like
$\Lambda N \to NN$, is not confirmed yet.
Recently, Schumacher proposed to check the validity of the
$\Delta I = \onehalf$ rule in the nonmesonic decays of the s-shell
hypernuclei{\cite{Schu}}.
It is therefore important to clarify the mechanism of the
$\Delta I= {1\over 2}$ rule in the free hyperon decays and to study
whether the same mechanism restricts the nonmesonic decays to $\Delta
I= {1\over 2}$ as well.

\subsection{Effective Weak Hamiltonian}

The effective weak hamiltonian describing $\Delta S = 1$ processes
has been calculated by several authors{\cite{VSZ,Okun,Paschos}}.
It can be computed by
evaluating perturbative QCD corrections using the operator product
expansion and the renormalization group equation for the Wilson
coefficients.
It should be noted that new four-quark operators mix to the pure weak
$su \to du$ vertex and thus change the flavor (isospin) structure
significantly.
The effective weak hamiltonian at the renormalization scale $\mu^2$
($\le 1$ GeV$^2$) is given by{\cite{Paschos}}
\eq
  H_{eff}^{\Delta S=1}=
  -\frac{G_f}{\sqrt 2}\sum_{r=1,r\ne 4}^6K_rO_r
\eq
where the four-quark operators, $O_k$ ($k= 1$, 2, 3, 5 and 6) are
defined by{\cite{VSZ}}
\eq
 O_1 &=& (\bar d_{\alpha}s_{\alpha})_{V-A}
          (\bar u_{\beta}u_{\beta})_{V-A}
         -(\bar u_{\alpha}s_{\alpha})_{V-A}
           (\bar d_{\beta}u_{\beta})_{V-A}\\
  O_2 &=& (\bar d_{\alpha}s_{\alpha})_{V-A}
           (\bar u_{\beta}u_{\beta})_{V-A}
         +(\bar u_{\alpha}s_{\alpha})_{V-A}
           (\bar d_{\beta}u_{\beta})_{V-A}\nonumber\\
     &+& 2(\bar d_{\alpha}s_{\alpha})_{V-A}
         (\bar d_{\beta}d_{\beta})_{V-A}
        +2(\bar d_{\alpha}s_{\alpha})_{V-A}
         (\bar s_{\beta}s_{\beta})_{V-A}\\
  O_3
     &=& O_3(\Delta I={1\over 2}) + O_3(\Delta I={3\over 2}) \\
  O_3(&\Delta& I={1\over 2}) = {1\over 3}\left[
           (\bar d_{\alpha}s_{\alpha})_{V-A}
           (\bar u_{\beta}u_{\beta})_{V-A}
         + (\bar u_{\alpha}s_{\alpha})_{V-A}
           (\bar d_{\beta}u_{\beta})_{V-A}\right.\nonumber\\
     &+& \left.2 (\bar d_{\alpha}s_{\alpha})_{V-A}
         (\bar d_{\beta}d_{\beta})_{V-A}
        -3 (\bar d_{\alpha}s_{\alpha})_{V-A}
         (\bar s_{\beta}s_{\beta})_{V-A}\right]\nonumber\\
  O_3(&\Delta& I={3\over 2}) = {5\over 3}\left[
           (\bar d_{\alpha}s_{\alpha})_{V-A}
           (\bar u_{\beta}u_{\beta})_{V-A}
         + (\bar u_{\alpha}s_{\alpha})_{V-A}
           (\bar d_{\beta}u_{\beta})_{V-A}\right.\nonumber\\
     &-& \left. (\bar d_{\alpha}s_{\alpha})_{V-A}
         (\bar d_{\beta}d_{\beta})_{V-A}\right]\nonumber\\
  O_5 &=& (\bar d_{\alpha}s_{\alpha})_{V-A}
       (\bar u_{\beta}u_{\beta}+\bar d_{\beta}d_{\beta}
      + \bar s_{\beta}s_{\beta})_{V+A}\\
  O_6 &=& (\bar d_{\alpha}s_{\beta})_{V-A}
       (\bar u_{\beta}u_{\alpha}+\bar d_{\beta}d_{\alpha}
      + \bar s_{\beta}s_{\alpha})_{V+A}
\eq
with
\eq
(\bar u_{\alpha}s_{\alpha})_{V-A} \equiv (\bar u_{\alpha}
        \gamma^{\mu}(1-\gamma_5) \, s_{\alpha}) \quad\rm{\etc}
\eq
$\alpha$ and $\beta$ denote the color of quarks and the
color sum is always assumed.

Among these operators,  $O_3$ contains a part that induces the $\Delta
I={3\over 2}$ transition, $O_3(\Delta I={3\over 2})$, while the others
are purely $\Delta I=\onehalf$.
The coefficients $K_r$ can be
calculated by solving the renormalization group equation to
the one-loop QCD corrections.
It is important to note that
the QCD correction enhances the $O_1$ coefficient and thus the $\Delta I
= \onehalf$ component,
while the purely weak four-quark vertex $su\to du$  contains both the
$\Delta I = \onehalf$ and $\half(3)$ components{\cite{VSZ}}.
Later we will compare the results with and without $O_3(\Delta I= {3\over 2})$
in order to study the $\Delta I= {3\over 2}$
contribution.
In the present calculation, we employ a set of parameters from
ref.\cite{Paschos} with $\mu=0.24$ GeV, which is determined so as to
give $\alpha_s (\mu^2)=1 $ for $\Lambda_{QCD} = 0.1$ GeV.
The values of the coefficients $K_r$
used in the present calculation are given in Table 1.

\begin{table}[t]
\caption{
Strengths of the weak effective four-fermi vertices,
taken from ref.\protect\cite{Paschos}.  We use the version with flavor-number
dependent $\Lambda$ and $m_t=200$GeV. The values of the CKM matrix
elements are taken as the central values of those given in
ref.\protect\cite{PData}.}
\begin{center}
\begin{tabular}{cc|ccccc}
 \hline
 $\mu $ (GeV) & $\Lambda^{(4)} $ (GeV) & $K_1$ & $K_2$ & $K_3$ &  $K_5$  &
 $K_6$\\
 \hline
 0.24& 0.10& $-0.284$ & $0.009$  & 0.026 & 0.004& $-0.021$\\
 \hline
\end{tabular}
\end{center}
\end{table}

This effective hamiltonian has been used for the calculations of the
nonleptonic decay of strange mesons and baryons{\cite{Okun}}.
It is found that although the $\Delta I= \onehalf$ transition is
indeed enhanced in those  decays,  the enhancement is not enough to
account for the experimental data quantitatively.
It was suggested{\cite{Sanda}} that an additional $\Delta I=\onehalf$
enhancement  arises from the mesonic correction in the
chiral effective theory.  We have confirmed their conclusion by using
the Nambu-Jona-Lasinio model without bosonization{\cite{Takizawa}}.

\subsection{The $\Lam N \to NN$ transitions\protect\footnotemark}
\footnotetext{This section is
based on the research carried out by T.~Inoue, S.~Takeuchi and
M.~Oka.\cite{ISO}}

The effective weak hamiltonian for quarks can be applied to
the valence quark model of two-baryon systems or six-quark systems.
We consider the direct quark (DQ) processes, \ie, the short-distance
weak interaction of two-baryons without mediated by mesons.
In calculating the DQ amplitudes for $\Lambda N \to
NN$, we employ
the constituent quark model, which describes the spin-flavor structure
of the ground-state baryons fairly well.
Two-baryon systems are expressed by
the quark-cluster-model wave functions{\cite{OY80,QCMYN}}:
\eq
\vert \Lambda N\rangle
                      &= &{\cal A}^6\vert \phi(1,2,3)\phi(4,5,6)
                          \chi_i(\vec{R})\rangle\nonumber\\
\vert NN\rangle
               &=&{\cal A}^6\vert \phi(1,2,3)\phi(4,5,6)
               \chi_f(\vec{R})\rangle
\label{eq:WF}
\eq
where ${\cal A}^6$ is the antisymmetrization operator for six quarks,
$\phi$ is the internal wave function of the baryon, and
$\vec{R}$ is the relative coordinate of two baryons.
$\chi_i(\Rv)$ ($\chi_f(\Rv)$)  is the initial (final)
relative wave function.
We assume that the orbital part of $\phi$ is a Gaussian with the
extension parameter $b=0.5$ fm, and the flavor-spin part of $\phi$ is
purely  SU(6) symmetric.

We here concentrate on the s-shell hypernuclei and assume that the
initial state is represented purely by $L=0$ and its wave function is
given by
\eq
\chi_i(\vec{R})=N_{0} g(\vec R)
                      \exp \left\{-\frac{1}{2B^2}\vec{R}^2\right\}
\eq
The final wave function is taken as
\eq
\chi_f(\vec R;\vec K )= g(\vec R)
                      \exp \left\{i\vec{K}\cdot\vec R \right\}
\eq
where $g$ represents the short range correlation,
\eq
g(\vec R) =  1 - C \exp\left[ - \frac{R^2}{r_0^2} \right]
\eq
We choose the parameter $B$ as $\sqrt2 \times 1.3$ fm
for the S-shell hypernuclei, which corresponds
to the shell model wave function with the Gaussian
parameter 1.3 fm for both the nucleon and $\Lambda$.
For the short range correlation we choose $ C=0.5 $ and $ r_0 = 0.5 $ fm
in the present calculation.
The relative momentum $K$ of the final state is determined by the
realistic $Q$ value $\Delta E\equiv M_{\Lambda} - M_N$ of the decay:
$K=416$ MeV/c.

The results of the calculation are summarized in Table 2.
As we restrict the initial state to $L=0$,
nine amplitudes, $a - f$ give all the information for the $\Lambda
N\to NN$ weak decay.
Amplitudes, $a$, $b$, and $f$ have contributions both from $\Delta
I=1/2$ and $\Delta I=3/2$ transitions.
The ratios $a_n/a_p$, $b_n/b_p$, and $f_n/f_p$ for $\Delta I=1/2$
are equal to $\sqrt 2$.
The other amplitudes, $c$, $d$ and $e$ do not contain any $\Delta I=
{3\over 2}$ component, because the final $NN$ states have $I=0$.
We find that the amplitudes $a$ and $b$ get large
$\Delta I= {3\over 2}$ contributions, while
$f$ is dominated by  $\Delta I= \onehalf$.
Thus we conclude that the $\Delta I= {3\over 2}$ transition can be
studied in the weak decay starting from the $\Lambda N$ $^1S_0$ state.

\begin{table}[t]
\caption{Calculated transition amplitudes in $10^{-10}$ $MeV^{-1/2}$.
This table is taken from ref.\protect\cite{ISO}}
%\label{tbl:amp}
\begin{center}
\begin{tabular}{lr||rrr||r}
\hline
 & & \multicolumn{3}{c||}{Direct Quark(DQ)} & OPE \\
 \multicolumn{2}{c||}{channel} & full & $\Delta I=1/2$ & $\Delta I=3/2$ & \\
\noalign{\hrule}
\multicolumn{2}{c||}{$p\Lambda\to pn$}&&&&   \\
$a_p$
 & ${}^{1}S_0\to {}^{1}S_0 $
& $-78.1$ & $ -23.4$ & $-54.7$ & $  2.2$ \\
$b_p$
 &          $\to {}^{3}P_0 $
& $-53.5$ & $  2.0 $ & $-55.5$ & $-24.8$ \\
$c_p$
               & ${}^{3}S_1\to {}^{3}S_1 $
& $ -1.0$ & $ -1.0$ & $  0  $ & $  2.2$ \\
$d_p$
                  &          $\to {}^{3}D_1 $
& $    0$ &     $0$ & $  0  $ & $-86.8$ \\
$e_p$
                  &          $\to {}^{1}P_1 $
& $-23.2$ & $-23.2$ & $  0  $ & $-43.0$ \\
$f_p$
                  &          $\to {}^{3}P_1 $
& $-55.4$ & $-53.8$ & $ -1.5$ & $ 20.2$ \\
\noalign{\hrule}
\multicolumn{2}{c||}{$n\Lambda\to nn$}&&&&   \\
$a_n$
 & ${}^{1}S_0\to {}^{1}S_0 $
& $ 5.5 $ & $ -33.1$ & $ 38.6$ & $  3.1$ \\
$b_n$
                  &          $\to {}^{3}P_0 $
& $42.2 $ & $  2.9$ & $ 39.3$ & $-35.1$ \\
$f_n$
                  & ${}^{3}S_1\to {}^{3}P_1 $
& $-75.1$ & $-76.2$ & $  1.0$ & $ 28.6$ \\
\noalign{\hrule}
\end{tabular}
\end{center}
\end{table}
\begin{table}[t]
\caption{Decay rates of light hypernuclei.
         All the decay rates are in the unit of  $\Gamma_{free}$.
         The experimental data for $\Gamma_{\Lambda p \to pn}$,
         $\Gamma_{\Lambda n \to nn}$, $\Gamma_{nm}(^5_{\Lambda}\mbox{He})$
         and $R_{np}({}^5_{\Lambda}\mbox{He})$ are taken from
         ref.\protect\cite{Szymanski}.
         Those for $R_{np}({}^4_{\Lambda}$He),
                   ${\Gamma_{nm}({}^4_{\Lambda}\mbox{He})}/
                          {\Gamma_{nm}({}^4_{\Lambda}\mbox{H}) }$
         and $\Gamma_{n0}/\Gamma_{p0}$ are taken from
	ref.\protect\cite{Schu2}.  This table is taken from
	ref.\protect\cite{ISO}.}
%\label{tbl:lnnn}
\begin{center}
\small
\begin{tabular}{l||cc||c||cc||c}
              & \multicolumn{2}{c||}{Direct Quark} & OPE
              & \multicolumn{2}{c||}{DQ $\pm$ OPE} & Exp
              \\
              & {Full}     & {$\Delta I=\frac12$ } &
              & $+$ &  $-$    &     \\
\noalign{\hrule}
$\Gamma_{p0} $  &  0.177  &  0.010  & 0.012
                &  0.235  &  0.143  & \\ % $0     \sim 0.116$   \\
$\Gamma_{p1} $  &  0.071  &  0.067  & 0.193
                &  0.260  &  0.269  & \\ %$0.074 \sim 0.187$   \\
$\Gamma_{n0} $  &  0.035  &  0.021  & 0.024
                &  0.002  &  0.118  & \\ %$0.063 \sim 0.553$   \\
$\Gamma_{n1} $  &  0.111  &  0.114  & 0.016
                &  0.042  &  0.212  & \\ %$0.049 \sim 0.196$   \\
\noalign{\hrule}
$\Gamma_{\Lambda p \to pn}$
                &  0.097  &  0.053  & 0.143
                &  0.253  &  0.238  & 0.105$\pm$0.035 \\
$\Gamma_{\Lambda n \to nn}$
                &  0.092  &  0.091  & 0.018
                &  0.032  &  0.189  & 0.100$\pm$0.055 \\
$\Gamma_{nm}\left( {}^5_{\Lambda}\mbox{He} \right)$
                &  0.378  &  0.295  & 0.333
                &  0.573  &  0.854  & 0.41 $\pm$ 0.14 \\
$R_{np}({}^5_{\Lambda}\mbox{He})$
                &  0.94   & 1.70    &  0.12
                &  0.12   & 0.79    &  0.93 $\pm$ 0.55\\
\noalign{\hrule}
$R_{np}({}^4_{\Lambda}$He)
                &  0.18   & 0.20    &  0.08
                &  0.004  & 0.24    &  0.18 $\pm$ 0.12 \\
$\frac{\Gamma_{nm}({}^4_{\Lambda}\mbox{He})}
      {\Gamma_{nm}({}^4_{\Lambda}\mbox{H}) }$
                &  0.63   & 0.66    &  6.58
                &  1.69   & 1.14    &  1.65 $\pm$ 0.77 \\
\noalign{\hrule}
$\Gamma_{n0}/\Gamma_{p0}$
                &  0.20   & 2.00    & 2.00
                &  0.01   & 0.854   & $-0.8 \pm 2.7$  \\
\noalign{\hrule}
$ a_1({}^5_{\Lambda}\mbox{He})$
                &  0.01   &  0.02   & $-0.19$
                &  0.20   & $-0.44$ &           \\
\noalign{\hrule}
\end{tabular}
\normalsize
\end{center}
\end{table}

The two-body transition rates are calculated for the $\Lambda N \to
nN$ ($N = n$ or $p$) transitions
with angular momentum $J$, and are denoted by  $\Gamma_{NJ}${\cite{BD}}.
Table 3 gives the results for $\Gamma_{NJ}$ as well as
the spin averaged  transition rates, defined by
\eq
   \Gamma_{\Lambda p\to pn}&=&
       \frac14\left( \Gamma_{p0} + 3\Gamma_{p1} \right)
   \ ,
   \\
   \Gamma_{\Lambda n\to nn}&=&
       \frac14\left( \Gamma_{n0} + 3\Gamma_{n1} \right)
\eq
We are also interested in the n--p ratio defined by
\eq
     R_{np}  \equiv
         \frac{\Gamma_{\mbox{neutron induced}}}
              {\Gamma_{\mbox{proton induced}}}
\eq
We find that the $J=0$ proton-induced transition rate, $\Gamma_{p0}$,
is strongly enhanced due to the $\Delta I=3/2$ transition.
The results of the one-pion exchange (OPE) transition are also shown in
Tables 2 and 3, for comparison{\cite{meson}}.
These amplitudes satisfy
$a_n/a_p=b_n/b_p=f_n/f_p=\sqrt2$,
because $\Delta I=\onehalf$ is assumed for the weak pion-baryon vertex.
One sees that the amplitude $d$ in this case is dominant.  This comes from
the tensor part of the one pion exchange and is enhanced due to a
large relative momentum in the final state.
Because the amplitude $d$ is not allowed for the $n\Lambda \to nn$ by
the Pauli principle, the $n-p$ ratio in the pion exchange amplitudes
becomes very small.
The DQ amplitude $d$ is neglected because the tensor part of DQ is
of order $(p/m)^2$.

Compared with the OPE result, $\Gamma_{p0}$ for DQ
is much larger and in fact is dominant while OPE is
dominated by the tensor transition included in $\Gamma_{p1}$.
This dominance of $\Gamma_{p1}$ in OPE makes the $n$-$p$ ratio,
$R_{np}$, small.
In DQ, $\Gamma_{n1}$ is also large so that the spin averaged $R_{np}$
is as large as 1.
Thus we find that DQ and OPE predict qualitatively different values
for $R_{np}({}^5_{\Lambda}\mbox{He})$, while they give similar
nonmesonic decay rate,
$\Gamma_{nm}(^5_{\Lambda}\mbox{He})$.
The experimental data prefers DQ, which indicates a significant
contribution of $\Gamma_{n1}$.
Because the pure $\Delta I =\onehalf$ calculation also yields large
$R_{np}$, its enhancement is not related to the $\Delta I=\onehalf$
rule violation.

Recently, Schumacher{\cite{Schu}} suggested that non-mesonic decays of
the A=4 and 5 hypernuclei may be useful in checking
the validity of the $\Delta I=\onehalf$ rule.
One can parametrize the n-p ratios of the
decay of $^4_{\Lambda}$He and $^5_{\Lambda}$He
and the ratio of non-mesonic decay widths of
$^4_{\Lambda}$He and $^4_{\Lambda}$H in terms of
$\Gamma_{NJ}${\cite{BD}}.
Then, using experimental data, the ratios of
$\Gamma_{NJ}$ can be extracted.
The $\Gamma_{n0}/\Gamma_{p0}$ ratio will be 2 for the pure
$\Delta I=\onehalf$ transition, while it becomes 1/2 for the pure
$\Delta I=\half(3)$ transition.
Our full amplitudes indeed predict small $\Gamma_{n0}/\Gamma_{p0}$,
which indicates a large $\Delta I=\half(3)$ contribution,
while the present
experimental data do not have enough precision to be conclusive.
It could be the first clear evidence for the $\Delta I=\half(3)$ weak
transition that is expected in the standard theory.

So far we have not considered the interference of the DQ and OPE
amplitudes.
The present formalism allows us to regard
OPE independent from DQ and therefore to superpose these two amplitudes.
Because the relative phase of two amplitudes are not known,
we evaluate DQ $\pm$ OPE and the results are given
in Table 3.
One finds that the difference between the two choices of the relative
phase mostly appear in the neutron-induced decay rates.
$\Gamma_{nJ}$'s are suppressed in (DQ $+$ OPE) and thus the $n-p$
ratio $R_{np}$ becomes very small.
In this sense, the experimental data prefer the (DQ $-$ OPE)
combination.
The ratio $\Gamma_{n0}/\Gamma_{p0}$ tends to be small ($ \ll 2$)
for both ( DQ $\pm$ OPE ) and again indicates a large $\Delta I =3/2$
contribution.
In both (DQ $\pm$ OPE), we find that $\Gamma_{\Lambda p \to pn}$
is overestimated and therefore the total nonmesonic decay rate
$\Gamma_{nm}({}^5_{\Lambda}\mbox{He})$ is too large. This is again due to
the large tensor component in $\Gamma_{p1}$ (OPE).
It seems important that OPE is calculated in the quark interaction
point of view in order to make a reliable prediction for the
DQ $-$ OPE interference.

\section{Conclusion}

I have reviewed several hadronic phenomena involving strangeness.  I
am particularly interested in the roles of the instanton induced
interaction in the hadron and multibaryon spectra, the YN interactions
from the quark model point of view and the strangeness changing weak
interaction among quarks.   These subjects are studied in simple quark
models, which give intuitive pictures of the underlying physics and
further give quantitative understanding of experimental data.   The
study of strange hadron systems is still far to go for complete
understanding and further effort both in experiment and theory is
anticipated. Present experimental facilities have already achieved a
significant progress in hypernuclear physics and are expected more
results. New facilities such as CEBAF and RHIC
are also expected to  shed new light on the strangeness physics.

\end{document}